\documentclass[modern]{aastex631}

    \usepackage{comment}
    \usepackage{outlines}
    \usepackage{amssymb}
    \usepackage{amsthm}
    \usepackage{amsmath}
    \usepackage{upgreek}
    \usepackage[fleqn]{mathtools}
    \usepackage[symbol]{footmisc}

    \newcommand{\bB}{ \mathbf{B} }

    \newcommand{\bE}{ \mathbf{E} }

    \newcommand{\bR}{ \mathbf{R} }
    \newcommand{\bS}{ \mathbf{S} }

    \newcommand{\bu}{ \mathbf{u} }

    \newcommand{\sA}{ {\scriptscriptstyle{\rm A}} }
    \newcommand{\sB}{ {\scriptscriptstyle{\rm B}} }

    \newcommand{\sM}{ {\scriptscriptstyle{\rm M}} }

    \newcommand{\sR}{ {\scriptscriptstyle{\rm R}} }
    \newcommand{\sS}{ {\scriptscriptstyle{\rm S}} }
    \newcommand{\sT}{ {\scriptscriptstyle{\rm T}} }

    \newcommand{\sSun}{ {\scriptscriptstyle{\rm \odot}} }

    \newcommand{\snull}{ {\scriptscriptstyle{\rm 0}} }
    \newcommand{\sone}{ {\scriptscriptstyle{\rm 1}} }
    \newcommand{\stwo}{ {\scriptscriptstyle{\rm 2}} }

    \newcommand{\muo}{ \mu_\snull }

    \newcommand{\bn}{\mathbf{\hat{n}} }

    \newcommand{\unI}{\mathbf{\hat{I}}}

    \newcommand{\grad}{ \overline{\mathbf{\nabla}} }
    \newcommand{\divgB}{ \overline{\mathbf{\nabla}} \cdot \bB }
    \newcommand{\divg}{ \overline{\mathbf{\nabla}} \cdot }
    \newcommand{\curl}{ \overline{\mathbf{\nabla}} \times }

    \newcommand{\cf}{{cf.,}}
    \newcommand{\eg}{{e.g.,}}

    \newcommand{\alf}{Alfv\'en }
    
    \newcommand{\amp}{Amp{\`e}re }
    \newcommand{\equationname}{Eq.}
    \renewcommand{\figurename}{Fig.}
    \renewcommand{\tablename}{Table}

\shorttitle{Stream-Aligned MHD}
\shortauthors{Sokolov et al.}

\graphicspath{{./}{figures/}}
\begin{document}

\title{Stream-Aligned Magnetohydrodynamics for Solar Wind Simulations}
\author[0000-0002-6118-0469]{Igor V. Sokolov}
\affiliation{Department of Climate and Space Sciences and Engineering\\
University of Michigan\\
2455 Hayward\\
Ann Arbor, MI 48109, USA}

\author[0000-0003-3936-5288]{Lulu Zhao}
\affiliation{Department of Climate and Space Sciences and Engineering\\
University of Michigan\\
2455 Hayward\\
Ann Arbor, MI 48109, USA}

\author[0000-0001-9360-4951]{Tamas I Gombosi}
\affiliation{Department of Climate and Space Sciences and Engineering\\
University of Michigan\\
2455 Hayward\\
Ann Arbor, MI 48109, USA}
\correspondingauthor{Igor Sokolov}
\email{igorsok@umich.edu}
\begin{abstract}
We present a reduced magnetohydrodynamic (MHD) mathematical model describing the dynamical behavior of highly conducting plasmas with frozen-in magnetic fields, constrained by the assumption that, there exists a frame of reference, where the magnetic field vector, $\mathbf{B}$, is  aligned with the plasma velocity vector, $\mathbf{u}$, at each point. We call this solution ``stream-aligned MHD'' (SA-MHD). Within the framework of this model, the electric field, $\mathbf{E} = -\mathbf{u} \times \mathbf{B} \equiv 0$, in the induction equation vanishes identically and so does the electromagnetic energy flux (Poynting flux), $\mathbf{E}\times\mathbf{B}\equiv0$, in the energy equation. 
At the same time, the force effect from the magnetic field on the plasma motion (the Amp{\`e}re force) is fully taken into account in the momentum equation. Any steady-state solution of the proposed model is a legitimate solution of the full MHD system of equations. However, the converse statement is not true: in an arbitrary steady-state magnetic field the electric field does not have to vanish identically (its curl has to, though). Specifically, realistic three-dimensional solutions for the steady-state (``ambient'') solar atmosphere in the form of so-called Parker spirals, can be efficiently generated within the stream-aligned MHD (SA-MHD) with no loss in generality.
\end{abstract}


\section{Introduction} \label{sec:intro}
\textit{Space weather} describes the dynamic state of the Earth's
magnetosphere-ionosphere system, which is driven by the solar wind and solar ionizing radiation.  The greatest disturbances in space weather are geomagnetic storms, the most severe of which are caused by coronal mass ejections (CMEs) (see \cite[]{Gosling:1993a}).
To simulate, forecast or just nowcast space weather events, which, from a mathematical standpoint, are essentially time-dependent processes, global models are needed to simulate, as the initial stage, the steady state of the solar-terrestrial environments. The resulting solution for the ``pre-event'' solar wind and terrestrial magnetosphere serves as the background through which the space weather disturbance propagates and shocks and discontinuities are generated. The solar energetic particle (SEP) transport as well as the formation of their seed population is mostly controlled by this background solution, as well. 

In realistic solar atmosphere simulations, the -- essentially three-dimensional (3-D) -- interplanetary magnetic field can be steady-state only in coordinate systems co-rotating with the Sun (such as HGR). In corotating frames solar active regions and coronal holes, that are mostly responsible for shaping the structure of the solar atmosphere, can be treated as steady-state sources. Despite being steady-state globally, the solution for the solar wind is highly variable at Earth's location, since in any corotating systems the Earth orbits the Sun with the \textit{Carrington rotation} period. This relative motion allows us to compare time series of solar wind parameters observed by Earth or L1 orbiting spacecraft, $SW_{observed}(t)$, with the simulated steady-state parameters for the 3-D background solar wind, $SW_{background}(\bR)$, by extracting the time series of simulated values, $SW_{background}(\bR_{Earth\_HGR}(t))$, at the time-dependent Earth location in the HGR system, $\bR_{Earth\_HGR}(t)$. The time series of observations by the two STEREO spacecraft provide additional validation possibilities.

In the last two decades several 3-D solar wind \cite[]{Usmanov:2000a, Suzuki:2005a, Verdini:2009a, Osman:2011a, Lionello:2014a, Lionello:2014b,Usmanov:2018a}, and coronal heating models \cite[]{tu97, hu00, dmit02, Habb03, cran10} that included, or were  exclusively driven by, \alf wave turbulence became increasingly popular. Our group also includes \alf wave turbulence in our corona and inner heliosphere model \cite[\cf][]{Sokolov:2013a, Sokolov:2021a, Oran:2013a, vanderHolst:2014a, Gombosi:2017a, Gombosi:2021rev}. Although popular, this physics-based approach to modeling the background solar wind is not the only way to model the solar corona and the solar wind. Empirical descriptions of the solar wind, like the widely used Wang-Sheeley-Arge (WSA) model \cite[]{Arge:2000a} are also attractive because of their simplicity and ability to predict the solar wind speed in the inner heliosphere. In addition, the WSA formalism can be readily incorporated into global 3-D models for the solar corona and inner heliosphere, e.g.,   via a varying polytropic index distribution \cite[see ][]{Cohen:2007a}, as proposed by \cite{Roussev:2003a}, or via the boundary condition set at the  solar surface \citep{Shen:2018a} or at $0.1$ AU  \cite[see, e.g.][]{Pomoell:2018a}. 

There is, however, a conceptual difference between global solar atmosphere models that are based on numerical solutions of the ideal (or resistive) MHD equations and simple (but efficient) semi-analytic solar wind models that imply a classical \textit{Parker-spiral} structure for the interplanetary magnetic field lines. Specifically, in semi-analytic models it is assumed that in the corotating frame of reference under steady-state conditions the solar wind velocity vector $\bu(\bR)$ is always aligned (parallel or anti-parallel) with the interplanetary magnetic field vector, $\bB(\bR)$. This means that stream-lines and magnetic field lines \textit{are always identical} and they form Archimedean or \cite{Parker:1958a} spirals. The assumed coincidence of solar wind stream-lines and interplanetary magnetic field lines (in corotating frames) is at the heart of almost all analytic, semi-analytic, semi-empirical and empirical models of interplanetary space. This assumption results in simple methods for tracing interplanetary magnetic field lines from any point of interest in the solar system back to the Sun. One can just trace back a Parker spiral to find out where the local solar wind and magnetic field line originates from on the Sun. For instance, in the WSA semi-empirical model \cite[]{Arge:2000a} this assumption is used to relate the Earth's location to the point on the \textit{source surface} at which to extract the parameters, predicted by the model. 

The problem with the concept of aligned interplanetary stream-lines and magnetic field lines is that numerical solutions of the full set of MHD equations have not been able to reproduce this feature so far. Full MHD numerical solutions of the solar atmosphere obtained within the framework of regular MHD are not stream-aligned. Even if the initial conditions describe parallel $\bB$ and $\bu$ vectors, the smallest numerical error (such as truncation, finite representation, etc) will result in an uncontrollable growth of misalignment. One of the reasons for this discrepancy is the numerical reconnection across the heliospheric current sheet near the top of helmet streamers: the reconnected field is directed across the current sheet, while the global solar wind streams along the current sheet, thus resulting in significant misalignment. Within regular MHD there is no mechanism to re-establish the stream-line-field line alignment.

Tracing magnetic field lines from a point of interest in the heliosphere to the solar surface ({\it magnetic connectivity}) is needed to identify the point of origin of the magnetic field line. This same magnetic connectivity is also necessary to solve the field-aligned transport of energetic particles that create space radiation hazards. With standard computational MHD calculating the magnetic connectivity can become so challenging that sometimes it is preferable to trace stream-lines instead, or even Lagrangian trajectories of fluid elements.

In this paper we present a new global model of the solar atmosphere/solar wind system 
using a \textit{reduced} MHD model which ensures stream-aligned magnetic field under steady-state conditions. We call this solution ``stream-aligned MHD'' (SA-MHD). In this reduced set of MHD equations the electric field, $\bE = - \bu \times \bB \equiv 0$, identically vanishes in the induction equation and so does the electromagnetic energy flux (Poynting flux) in the energy equation, $\bE \times \bB \equiv 0$. At the same time the impact of the magnetic force on the plasma motion (the \amp force) is explicitly enforced in the momentum equation. It must be emphasized that any steady-state solution of the reduced equation set is a \textit{solution} of the full MHD system of equations. However, the converse statement is not true, since in an arbitrary steady-state magnetic field the electric field does not have to vanish identically (though its curl must). The main point is that the practically interesting \cite{Parker:1958a} spiral solutions for the steady-state background solar atmosphere and solar wind can be efficiently generated within the SA-MHD with no loss of generality, in 3-D geometry with realistic solar magnetic field.
\section{Full 3-D MHD Equations of the Global Model}
\subsection{Standard MHD Equations with Non-Zero Magnetic Field Divergence}
Global models of the solar atmosphere 
are usually based on the standard MHD equations (non-specified notations are as usual): 
\begin{subequations}
\begin{align}
\label{eq:continuity}
& \partial_t\rho+\divg(\rho{\bu})=0,
\\
\label{eq:induction}
& \partial_t\bB+\divg\left(\bu\bB-\bB\bu\right)= - \bu(\divgB),
\\
\label{eq:momentum}
&\partial_t(\rho\bu)+\divg\left[\rho\bu\bu-\frac{\bB\bB}{\muo}+\left(P+\frac{B^2}{2\muo}\right)\unI\right]=-\frac1{\muo}\bB(\divgB),
\\
\label{eq:totenergy}
&\partial_t\left(\frac{\rho u^2}2+\frac{P}{\gamma-1}+\frac{{\bf B}^2}{2\mu_\snull}\right) + \grad\cdot\left[\left(\frac{\rho u^2}2+\frac{\gamma P}{\gamma-1}\right){\bf
  u}\right]
  \nonumber\\ & \qquad
  +\grad\cdot\left[
  \frac{\left(\bu\bB-\bB\bu\right)\cdot\bB}{\mu_\snull}\right] = -\frac{({\bf u}\cdot\mathbf{B})}{\mu_\snull}(\grad\cdot\mathbf{B}),
\end{align}
\end{subequations}
where $B=|\bB|$, $u=|\bu|$, and $\unI$ is the unit tensor. We use the equation of state, $P = 2n_i k_\sB T$, $\rho=m_pn_i$ for the coronal plasma with the polytropic index, $\gamma=5/3$. We omitted many important effects \cite[effects of gravity and inertial forces are discussed in section~\ref{sec:rotframe}, see][for detailed description of a model used in simulations here]{Sokolov:2021a} in the governing equations, while we explicitly included terms proportional to the divergence of magnetic field. If the numerical solution for the magnetic field is not divergence-free \cite[see][]{Godunov:1972a, Powell:1999a}, this approach allows us to handle maintain basic physical properties, such as pressure positivity. Indeed, if one takes \equationname~(\ref{eq:totenergy}) divided by temperature, $T$, subtracts the dot product of $\bu/T$ with \equationname~(\ref{eq:momentum}) and the dot product of $\bB/(\muo T)$ and \equationname~(\ref{eq:induction}) and finally adds \equationname~(\ref{eq:continuity}) times  $s+\frac1T\left(\frac{u^2}{2}-h\right)$, an extra conservation law is obtained:
\begin{equation}
\label{eq:entropy}
    \partial_t(\rho s) + \divg(\rho\bu s)=0
\end{equation} 
\cite[see also][]{Godunov:1961b, Godunov:1972a, Harten:1983a, Powell:1999a}, where $s$ and $h$ are the entropy and enthalpy per unit mass and the thermodynamic equations
\begin{equation}
\label{eq:thermo}
  dh=  T ds 
    + \frac1\rho  dP,\qquad h=\frac{\gamma P}{\rho(\gamma - 1)},
\end{equation}
were used. \equationname~(\ref{eq:entropy}) is only valid for smooth solutions and it states that the entropy satisfies the conservation law for a passively advected scalar. At shocks and discontinuities the entropy conservation breaks down, but, entropy can only increase, not decrease. Therefore, the total entropy (its volume integral) does not decrease:
    $\partial_t\int{\rho s \mathrm{d}V} \ge 0$,
since the integral of $\partial_t(\rho s)$ over the smooth solution region vanishes due to Gauss' theorem applied to Eq.~(\ref{eq:entropy}), while the discontinuities result in entropy production. A consequence of this entropy property is that for such systems finite-volume numerical schemes ensure that in the numerical solution the physical entropy does not decrease, and consequently, the pressure can never become negative (should the pressure go to zero, the entropy, $s\propto\log(P)$, would drop to minus infinity).

\subsection{Separating the Potential Magnetic Field in the Solar Corona}
Realistic models of the 3-D coronal magnetic field use boundary conditions taken from full disc photospheric magnetograms incorporating current and past observation results. The potential magnetic field solution provides the minimum free magnetic energy for a given boundary condition, therefore, in the ``ambient'' (background) solution for the solar wind the magnetic field is approximately equal to the potential configuration in the proximity of the Sun. Following \cite{Ogino:1984a} and \cite{Tanaka:1994a} \cite[see also][]{Powell:1999a, Gombosi:2002a} we split the total magnetic field to a potential ($\bB_\snull$) and non-potential ($\bB_\sone$) component $\bB=\bB_\snull +\bB_\sone$. We note that at  $R=R_\sSun$ the \textit{potential} ${\bf B}_\snull$ field dominates. If we use photospheric maps of the radial magnetic field ($B_r$), then the potential $\bB_\snull$ field can be recovered from the observed magnetogram using the Potential Field Source Surface Method (PFSSM) that has been originally described by \cite{alt77}. The Laplace \equationname~for scalar magnetic potential is solved in terms of spherical harmonics between $R_\sSun \leq R$ and $R \leq R_{\sS\sS} = 2.5R_\sSun$ with the given radial gradient of the potential (the observed radial field) at $R=R_\sSun$ and with vanishing magnetic potential (i.e. purely radial magnetic field) at $R=R_{\sS\sS}$. Accordingly, the non-potential ($\bB_\sone$) field (used by \cite{Sokolov:2013a, Oran:2013a, vanderHolst:2014a}) satisfies zero boundary condition for the radial field component, $\bB_{\sone_R} = 0$ at $R = R_\sSun$. 

Using the split magnetic field approach the governing equations become the following (the continuity equation does not change, therefore we do not repeat it here):
\begin{subequations}
\begin{align}
\label{eq:inductionB0}
& \partial_t\bB_\sone  +\divg\left[\bu(\bB_\sone+\bB_\snull)-(\bB_\sone+\bB_\snull)\bu\right]= - \bu\left[\divg(\bB_\sone+\bB_\snull)\right],
\end{align}
\begin{align}
\label{eq:momentumB0}
& \partial_t(\rho\bu)  + \divg\left[\rho\bu\bu-\frac{\bB_\sone\bB_\sone+\bB_\snull\bB_\sone + \bB_\sone\bB_\snull}{\muo} + \left(P+\frac{B_\sone^2}{2\muo} + \frac{\bB_\sone\cdot\bB_\snull}{\muo}\right)\unI\right]
\nonumber\\
& \quad  =  -\frac{\bB_\sone}{\muo}\left[\divg(\bB_\sone+\bB_\snull)\right]
-\frac{\bB_\snull}{\muo}(\divg\bB_\sone)
-\frac1{\muo}[\grad\times\bB_\snull]\times\bB_\sone,
\end{align}
\begin{align}
\label{eq:totenergyB0}
& \partial_t \left(\frac{\rho u^2}2 + \frac{P}{\gamma-1} + \frac{ B_\sone^2}{2\mu_\snull}\right) + \divg\left[\left(\frac{\rho u^2}2+\frac{\gamma P}{\gamma-1}\right){\bf u}\right]
  \nonumber\\ 
& \quad + \divg\left[
  \frac{\left[\bu\left(\bB_\sone+\bB_\snull\right)-
  \left(\bB_\snull+\bB_\sone\right)\bu\right]\cdot\bB_\sone}{\mu_\snull}\right] = -\frac{(\bu\cdot\bB_\sone)}{\mu_\snull}\left[\divg(\bB_\sone+\bB_\snull)\right].
\end{align}
\end{subequations}

An advantage of splitting the magnetic field is that there are no quadratic terms in $\bB_\snull$ in equation~(\ref{eq:momentumB0}). Near the Sun (especially near active regions) such terms can be so large, that the error in their approximation might exceed the whole physical effect from the $\bB_\sone$ field. For the same reason we cannot, in general, rely on the assumption that numerical approximations for $\curl\bB_\snull$ and $\divg\bB_\snull$ are exactly zero (as it is theoretically assumed), so that to maintain entropy conservation we need to introduce the corresponding source terms in equations~(\ref{eq:inductionB0}--\ref{eq:totenergyB0}). Particularly, all contributions from the field to the momentum \equationname~(\ref{eq:momentumB0}) can be combined with the \amp force $\mathbf{j}\times(\bB_\sone + \bB_\snull) = \frac1{\muo}[\grad\times\bB_\sone]\times(\bB_\sone + \bB_\snull)$ and it is the last term in the RHS of (\ref{eq:momentumB0}), which ensures that the potential, $\bB_\snull$, contributes to the total field, but not to the current, in the \amp force.  

In a solar corona model different approximations, with split and non-split magnetic fields, can be used depending on the heliocentric distance, $R=|\bR|$. For instance, the ``threaded field line model'' \cite[]{Sokolov:2021a} that includes the transition region and covers heliocentric distances between the photosphere ($R_\sSun$) and  the low boundary of the solar corona, $R_b\approx 1.1R_\sSun$, neglects the non-potential ${\bB}_\sone$ component, assuming that in this region ${\bB}_\sone\equiv0$, and the plasma flow occurs only along the potential field lines (\textit{threads}). This approach allows us to move the boundary conditions for all other physical quantities from the top of the transition region (where they are poorly known) to the photosphere. Accordingly, the boundary condition, $({\bB}_\sone)_R=0$, is imposed at $R=R_b$, to solve the full set of equations~(\ref{eq:inductionB0}--\ref{eq:totenergyB0}) at distances $R_b\le R\le R_{\sS\sS}$.

Finally, above the magnetogram source surface, at $R\ge R_{\sS\sS}$, the $\bB_\snull$ field, which is purely radial at $R_{\sS\sS}$, can be naturally continued beyond the source surface as a steady-state, radial and divergence-free field decaying as $1/R^2$:
\begin{equation}
\label{eq:B0source}   
\bB_\snull(\bR)=
\frac{R^2_{\sS\sS}}{R^2}
\bB_\snull\left(\frac{R_{\sS\sS}}{R}\bR\right).
\end{equation}
This magnetic field, however, is no longer a potential field, therefore, its contribution to the current
cannot be neglected. To fully account for the total force effect when using split fields and to ensure their continuity at $R_{\sS\sS}$, one needs to add the extra source term to the RHS of the momentum \equationname~(\ref{eq:momentumB0}) in the $R\ge R_{\sS\sS}$ region:
\begin{equation}
\label{eq:momentumsource}  
{\bf S}_\sM=\frac1{\mu_0}\divg\left(\bB_\snull\bB_\snull\right)-\frac1{2\mu_0}\grad B^2_\snull-\frac1{\mu_0}\bB_\snull(\divg\bB_\snull)+\frac1{\mu_0}[\grad\times\bB_\snull]\times\bB_\sone,
\end{equation}
which is easy to compute, since only the last term (which is to cancel the last term in the RHS of \ref{eq:momentumB0}) depends on time via $\bB_\sone$, while all other terms are steady-state terms and can be calculated only once. 

With the newly added source the momentum equation, in effect, reduces to \equationname~(\ref{eq:momentum}) expressed in terms of the total magnetic field, $\bB=\bB_\snull + \bB_\sone$ and the induction equation (\equationname~\ref{eq:inductionB0}) reduces to (\ref{eq:induction}). Nevertheless, it is the unknown field $\bB_\sone$ which is solved from these equations throughout computational domain. In order not to break the entropy conservation, for $R\ge R_{\sS\sS}$ the corresponding source term, $S_E=\bu\cdot{\bS}_\sM$, should be added to the RHS of the energy \equationname~(\ref{eq:totenergyB0}). For derivations, but not for computations, the RHS of resulting energy equation can be written as follows:
\begin{eqnarray}
\label{eq:energytotRHS}
    \bu\cdot{\bf S}_\sM & - & \frac{(\bu\cdot\bB_\sone)}{\mu_\snull} \left[\divg(\bB_\sone + \bB_\snull)\right] = -\frac{\left[\bu\cdot(\bB_\sone + \bB_\snull)\right]}{\mu_\snull} \left[\divg(\bB_\sone  \bB_\snull)\right] \nonumber \\
    & + & \frac1{\muo}\divg\left\{\left[(\bB_\sone + \bB_\snull)\bu - \bu(\bB_\sone + \bB_\snull)\right]\cdot\bB_\snull\right\} - \frac1{\muo}\bB_\snull\cdot\partial_t\bB_\sone.
\end{eqnarray}
Although the energy \equationname~(\ref{eq:totenergyB0}) with the split field and the transformed RHS is formally reducible to Eq.~(\ref{eq:totenergy}) for the full field, such reduction would require us to redefine the magnetic energy density in terms of the total field too, as $(\bB_\sone + \bB_\snull)^2/(2\muo)$, so that its time derivative, $(1/\muo) (\bB_\sone + \bB_\snull) \cdot \partial_t \bB_\sone$, balances the last term in the RHS of Eq.~(\ref{eq:energytotRHS}). This redefinition, however, may compromise the whole idea of only solving equations for $\bB_\sone$ beyond $R_{\sS\sS}$. To use the energy equation, (\ref{eq:totenergyB0}), with an extra source, $\bu\cdot{\bf S}_\sM$, seems to be a better option. 

\section{Reduced Equations of SA-MHD}
\subsection{Intuitive solution}
Now, we show how to proceed from standard MHD equations (\ref{eq:continuity}-\ref{eq:totenergy}) to equations for SA-MHD, in which
\begin{equation}
\label{eq:alignment}
    \bB = \alpha\,\bu,
\qquad
    \bE = -\bu \times \bB \equiv 0,
\end{equation}
where $\alpha(t,\bR)$ is a scalar function of time and coordinates. To derive a governing equation for $\alpha$, in a steady-state stream-aligned solution, one can integrate the MHD equations over a magnetic flux tube element of a length of $d\ell$ bounded by two nearby cross sections of the flux tube, $dS_\sone$ and $dS_\stwo$, and a bundle of magnetic field lines about a chosen magnetic field line, which all pass through the contours of these cross sections. The equation, $\divgB=0$, gives 
\begin{equation}
\label{eq:Bflux}
    BdS={\rm const}
\end{equation}
along the flux tube,  which allows us to relate the change in the cross-section area along the thread to the magnetic field magnitude. Now, since the stream-lines are aligned with the magnetic field lines, one can also integrate the continuity \equationname~(\ref{eq:continuity}) for a steady-state flow along the flux tube, which gives
\begin{equation}
\label{eq:Mflux}
    \rho u S={\rm const}
\end{equation} 
Therefore, the ratio of constant mass flux and the constant magnetic flux, \begin{equation}
\label{eq:ratio}
    \frac{B}{\rho u}\equiv \frac{\alpha}\rho
\end{equation}
is constant along the field line and along the stream-line. Both of these relations can be expressed as
\begin{equation}
\label{eq:field-stream}
    (\bu\cdot\grad)\frac{\alpha}{\rho}=0
\end{equation}
Now, if we assume that this ratio is constant along the fluid particle trajectory (which in steady state coincides with the stream-line but in the dynamic system this statement is a matter of a model assumption): 
\begin{equation}
\label{eq:conserve}
    \partial_t \left(\frac{\alpha}{\rho}\right) + (\bu\cdot\grad)\left(\frac{\alpha}{\rho}\right) = 0,
\end{equation}
we can combine equations (\ref{eq:continuity}) and (\ref{eq:conserve}) to obtain the conservation law for $\alpha$:
\begin{equation}
\label{eq:cglevolution}
    \partial_t \alpha = - \divg \left(\alpha\bu\right).
\end{equation}

\subsection{Rigorous solution}
A less intuitive, but mathematically more rigorous derivation of the same set of model equations can be obtained if we consider an MHD state that is close to the force equilibrium, $\partial_t\bu \rightarrow 0$, and add the \textit{aligning source} (that tends to zero) to the induction equation (\equationname~\ref{eq:induction}):
\begin{equation}
\label{eq:aligningsource1}
    \partial_t\bB + \divg \left(\bu\bB - \bB\bu\right) = - \bu(\divgB) + \alpha \partial_t\bu , \qquad
    \alpha = \frac{\bB\cdot\bu}{u^2},
\end{equation}
where the definition of $\alpha$ from Eq.~(\ref{eq:alignment}) is generalized for the case of non-aligned vectors $\mathbf{u}$ and $\mathbf{B}$; for the aligned vectors both definitions are identical. To keep the entropy conservation, a similar source should be added to the RHS of the energy (\equationname~\ref{eq:totenergy}):
\begin{align}
\label{eq:aligningsource2}
    & \partial_t \left( \frac{\rho \bu^2}2 + \frac{P}{\gamma-1} +\frac{\bB^2}{2\muo}\right) +
    \divg\left[\left(\frac{\rho \bu^2}{2} + \frac{\gamma P}{\gamma-1}\right) \bu \right] + 
    \nonumber \\ & \qquad +
    \divg\left[\frac{\bB^2\bu -\bB(\bu\cdot\bB)}{\mu_\snull}\right]
    = -\frac{(\bu\cdot\bB)}{\muo}(\grad\cdot\mathbf{B}) + \frac{\alpha}{\muo}\bB\cdot\partial_t\bu
\end{align}
Now, if in an arbitrary MHD initial state the magnetic field is aligned according to Eq. (\ref{eq:alignment}),  the aligning term ensures, that 
\begin{equation}
\label{eq:cglEfield}
    \partial_t\bE = \left[\partial_t\bB \times \bu\right] - \left[\partial_t\bu \times \bB\right] = \left[\partial_t\bu \times\left(\alpha\bu-\bB\right)\right] = 0,
\end{equation}
and keeps the alignment forever. By substituting $\bB=\alpha\bu$ into \equationname~(\ref{eq:aligningsource1}), we arrive at equation~(\ref{eq:cglevolution}) (multiplied by $\bu$). 

In the momentum equation the stream-aligned magnetic field should be applied:
\begin{equation}
\label{eq:cglmomentum}
    \partial_t(\rho\bu) + \divg\left[\rho \bu\bu + P \unI - \frac{\alpha^2}{\muo} \left(\bu\bu - \frac{u^2}{2}\unI\right)\right] = - \frac{\alpha}{\muo} \bu \divg\left(\alpha\bu\right),
\end{equation}
which can be also written in a non-conservative form:
\begin{equation}
\label{eq:cglmomentum2}
    \partial_t\bu + (\bu\cdot\grad)\bu = -\frac{\grad P}{\rho} + \frac{\alpha}{\muo\rho} \left[\curl\left(\alpha\bu\right)\right]\times\bu
\end{equation}
\equationname~(\ref{eq:cglmomentum2}) contains the explicit expression for the Amp{\`e}re force, which happens to be always perpendicular to the velocity vector (similarly to the Coriolis force). In the equation of energy (\ref{eq:totenergy}) the Poynting flux, $\bE\times\bB$ vanishes identically. In addition, the sources in the RHS of Eq.~(\ref{eq:aligningsource2}) for the stream-aligned field reduce to $\partial_t B^2/2\muo$ and entirely balance the change in the magnetic energy density in the LHS. Therefore in effect, the stream-aligned magnetic field does not contribute to the energy density, so that the energy equation does not include the magnetic field at all:  
\begin{equation}
\label{eq:cglenergy}
\partial_t \left(\frac{\rho u^2}2+\frac{P}{\gamma-1}\right)  +  \divg\left[\left(\frac{\rho u^2}2 + \frac{\gamma P}{\gamma-1}\right)\bu\right]=0.
\end{equation}
This is because the Amp{\`e}re force is perpendicular to the velocity vector and it does not affect the kinetic energy, hence, does not exchange energy between the field and plasma. For the same reason, under steady-state conditions, equation~(\ref{eq:cglenergy}) conserves the Bernoulli integral along the stream-line in its classical form:
\begin{equation}
\label{eq:bernoulli}
    (\bu\cdot\grad) \left[\frac{u^2}{2}+\frac{\gamma P}{(\gamma-1)\rho}\right]=0.
\end{equation}

Equations (\ref{eq:continuity},\ref{eq:cglevolution},\ref{eq:cglmomentum},\ref{eq:cglenergy}) present a full set of governing equations, which, as the solution approaches steady state, becomes fully conservative (the nonconservative source term that is proportional to $\partial_t\alpha$ in the momentum equation tends to zero), satisfies the $\divgB=0$, constrain (since, $\divgB=-\partial_t \alpha \rightarrow 0$), and their solution in this limit presents a legitimate solution of the MHD equations, since the aligning sources ($\propto\partial_t \bu\rightarrow0$) tend to zero too. Moreover, this is a desired stream-aligned solution. On the other hand the full MHD equations with no aligning sources, once applied to any non-equilibrium MHD state, with $\partial_t\bu\ne0$, will immediately produce non-stream-aligned solution even from stream-aligned initial state. It seems that in any discretized implementation (actual numerical solution) it is extremely difficult (if not impossible) to reach stream-aligned steady-state 3-D solutions within the framework of regular computational MHD, which justifies the approach proposed here.
\subsection{Stream Aligned MHD in a Rotating Frame of Reference}
\label{sec:rotframe}
For applications to solar wind the momentum and energy equations of SA-MHD are used the frame of reference rotating with angular velocity, ${\bf \Omega}_\sSun$, with respect to the inertial frame. In addition to the Coriolis force density, $-2\rho{\bf \Omega}_\sSun\times\bu$,  the centrifugal and gravitational forces are accounted for via the gradient of the potential, $\Phi(\bR)$:
\begin{align}
\label{eq:inertialpotential}
& \Phi(\bR) = -\frac{GM_\sSun}R - \frac12{\left[{\bf \Omega}_\sSun \times\bR\right]^2},
\\
\label{eq:rotmomentum}
& \partial_t(\rho\bu) + \divg\left[\rho \bu\bu + P \unI - \frac{\alpha^2}{\muo} \left(\bu\bu - \frac{u^2}{2}\unI\right)\right] = - \frac{\alpha}{\muo} \bu \divg\left(\alpha\bu\right)-2\rho{\bf \Omega}_\sSun\times\bu-\rho\grad\Phi,
\\
\label{eq:rotenergy}
& \partial_t \left[ \rho\left(\frac{u^2}2+\Phi\right)+\frac{P}{\gamma-1}\right] + \divg\left\{\left[\rho\left(\frac{ u^2}2+\Phi\right) + \frac{\gamma P}{\gamma-1}\right]\bu\right\}=0.
\end{align}
In the rotating frame the Bernoulli integral (\equationname~\ref{eq:bernoulli}) is modified by the inertial force potential (\equationname~\ref{eq:inertialpotential}):
\begin{equation}
\label{eq:rotbernoulli}
    (\bu\cdot\grad) \left[\frac{u^2}{2} - \frac{GM_\sSun}R - \frac12{\left[{\bf \Omega}_\sSun\times\bR\right]^2} + \frac{\gamma P}{(\gamma-1)\rho}\right]=0.
\end{equation}

The Bernoulli integral can be used to relate the model predictions at the source surface close to the Sun, such as the WSA predictions at $R=2.5\,R_\sSun$, to the observed solar wind parameters at 1 AU \cite[see \eg][]{cohen07}. However, \equationname~(\ref{eq:rotbernoulli}) is not convenient for use, since the velocity in the inertial frame, $\bu_{\rm inert}$, relates to that in the rotating frame, $\bu$, with the well-known equation:
\begin{equation}
\bu_{\rm inert}=\bu+{\bf \Omega}_\sSun\times\bR,
\end{equation}
so that at 1 AU the velocity in the rotating frame has too large contribution from rotational velocity. In terms of the observable parameters in the inertial  frame, the Bernoulli integral implies the conservation of the following quantity:
\begin{equation}
\label{eq:bernoulliinertial}
\frac{u^2_{\rm inert}}{2} - {\bf \Omega}_\sSun\cdot\left[\bR\times\bu_{\rm inert}\right] - \frac{GM_\sSun}R +h = {\rm const}, \qquad h=\frac{\gamma P}{(\gamma-1)\rho}.
\end{equation}
Within the framework of isothermal approximation for solar atmosphere, one can use generic expression enthalpy per a unit of mass, $h$, instead of a particular model for ideal gas with constant $\gamma$ assumed in \equationname~\ref{eq:bernoulliinertial}. For isothermal two-component plasma with $T_i=T_e=T$, the expression for enthalpy is $h=\frac{2k_BT}{m_p}\log\rho$. 

Note, that in neglecting the effect from the gas-kinetic pressure and from the magnetic field, 
for a test particle of a unit mass co-moving with the solar stream velocity, $\bu$, both the energy integral, $\frac{u^2_{\rm inert}}{2}- \frac{GM_\sSun}R={\rm const}$, and the vector of particle angular momentum,
$\left[\bR\times\bu_{\rm inert}\right]={\rm const}$, conserve separately and so does their linear combination in \equationname~\ref{eq:bernoulliinertial}. However, the force effect of stream-aligned magnetic field increases the angular momentum of the solar wind while moving outward the Sun. Therefore, according to \equationname~\ref{eq:bernoulliinertial} at large heliocentric distances there is a finite gain in the solar wind energy per unit of mass equal to ${\bf \Omega}_\sSun\cdot\left[\bR\times\bu_{\rm inert}\right]_{1\,{\rm AU}}$

The closed system of equations describing self-consistent variation of the angular momentum and azimuthal magnetic field component while the solar wind propagating outward the Sun  \cite[]{Parker:1958a} can be obtained by solving the equations of SA-MHD in spherical coordinates, $R,\theta,\varphi$, which are the heliocentric distance, co-latitude and longitude correspondingly. The analytical solution may be found assuming  no dependence on  $\varphi$ (axially symmetric field) and no motion over $\theta$ (radially divergent flow). The Coriolis force forces the particles to rotate in the negative $\varphi$ direction, while it has no component in the $\theta$ direction. Consequently, the fluid motion occurs only in the $(R,\varphi)$ direction and all stream-lines will be on conical surfaces of constant $\theta$. Steady-state stream-aligned solutions of Eqs. (\ref{eq:continuity}, \ref{eq:rotmomentum}, \ref{eq:rotenergy}) may be integrated along stream-lines yielding following equation:
\begin{equation}
\label{eq:parker2}
    R \sin\theta \left[\left(1-\frac{\alpha^2}{\muo\rho}\right) u_\varphi + \Omega_\sSun R \sin\theta\right] = \mathrm{const} = \Omega_\sSun \sin^2\theta\left\{R^2\right\}_{\sA}.
\end{equation}
In the absence the magnetic field, i.e.\ at $\alpha\equiv0$, \equationname~(\ref{eq:parker2}) would express the conservation of angular momentum ($R \sin\theta u_{\varphi,{\rm inert}}\equiv\left|\bR\times\bu_{\rm inert}\right|$) in the inertial (not rotating!) system.
However, due to the effect from realistic interplanetary magnetic field the heliocentric distance in the right hand side should be taken at the \alf point where the local stream speed, $u$, equals to the \alf wave speed, $V_\sA=B/\sqrt{\muo\rho}$, since
\begin{equation}\label{eq:inverseMA}
\frac{\alpha^2}{\muo\rho}=\frac{V_A^2}{u^2}=\frac{1}{M_\sA^2}.
\end{equation}

Typically, the factor $M_\sA^{-2}$ is very large near the base of the solar corona, because in this region the solar magnetic field is large and the solar wind not accelerated yet. In the opposite limiting case at large heliocentric distances, where the solar wind is fast and the magnetic field is weak, the plasma stream is hyperalfv{\'e}nic, $M_\sA^{-2}$ is negligibly small. In between the accelerating and rarefying solar wind must go through the \alf point and this is the point at which the integration constant is determined (this point to avoid discontinuous solutions). With these considerations the azimuthal velocity is \cite[see][]{Weber:1967a}:
\begin{equation}
\label{eq:parker3}
    u_\varphi=\frac{\Omega_\sSun\sin\theta \left[\left(R^2\right)_{\sA} - R^2\right]}{R  \left(1 - \frac{1}{M_\sA^2}\right)}.
\end{equation}
\begin{figure*}[htb]
\centering
\includegraphics[width=0.4\textwidth]{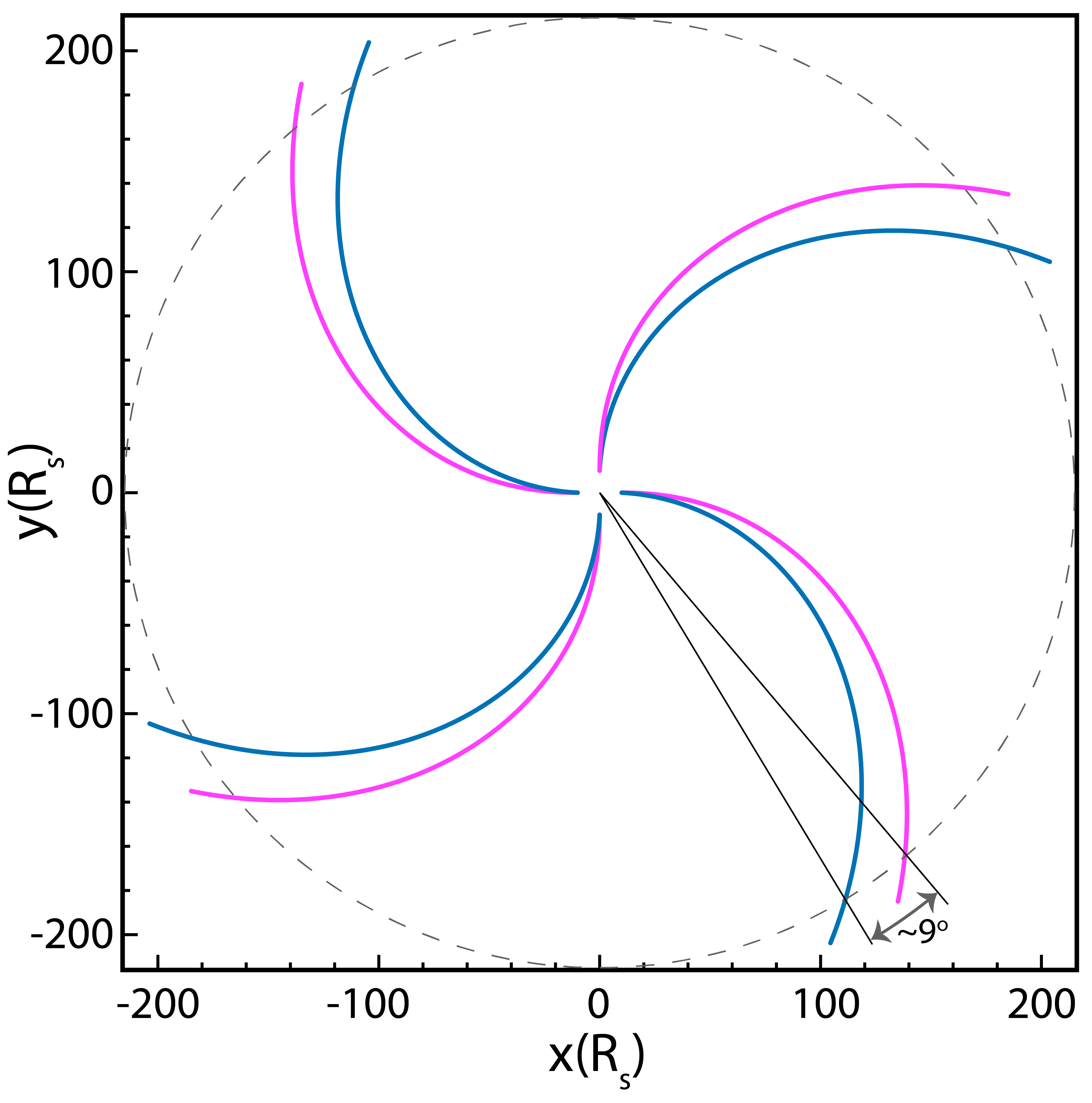}
{\caption{\cite{Parker:1958a} solutions (magenta) vs the SA-MHD solution by \cite{Weber:1967a} (blue) in the equatorial plane for a uniform solar corona. A constant solar wind speed of $u_\sR=400$ km/s and source surface radius of $10\,R_\sSun$ are assumed.
}
\label{fig:Parker}}
\end{figure*}

\figurename~\ref{fig:Parker} demonstrates the difference between the over-simplified \cite{Parker:1958a} spiral and the SA-MHD solution according to \equationname~(\ref{eq:parker3}). One can see that under idealized conditions (constant solar wind speed of $u_\sR=400$ km/s and source surface radius of $10\,R_\sSun$) the more accurate stream-aligned solution \equationname~(\ref{eq:parker3}) lags behind the \cite{Parker:1958a} spiral by about $9^\circ$  at Earth's orbit. Taking into account the solar rotation (with respect to Earth) this means that a stream-aligned field line originating from the same point of the solar surface will cross Earth about 15 hours later than the corresponding \cite{Parker:1958a} spiral aligned field line. In space weather applications, tracing the stream-line in the co-rotating frame from the observation point toward its origin site can be used to associate \textit{in situ} solar wind properties to remotely observed structures of the solar corona \cite[see, e.g.][]{Biondo2021a}. The large difference demonstrated in \figurename~\ref{fig:Parker}  means that field line associated phenomena (like SEP events) will arrive to Earth some 15 hours later than predicted by simple \cite{Parker:1958a} spiral connectivity models. It is, therefore, important that our computational model fully incorporates all principles and capabilities of earlier models.

\section{Simulation Results and Discussion}

\subsection{Numerical Solution with Full MHD}
\begin{figure}[h!t]
\centering
\includegraphics[width=1\textwidth]{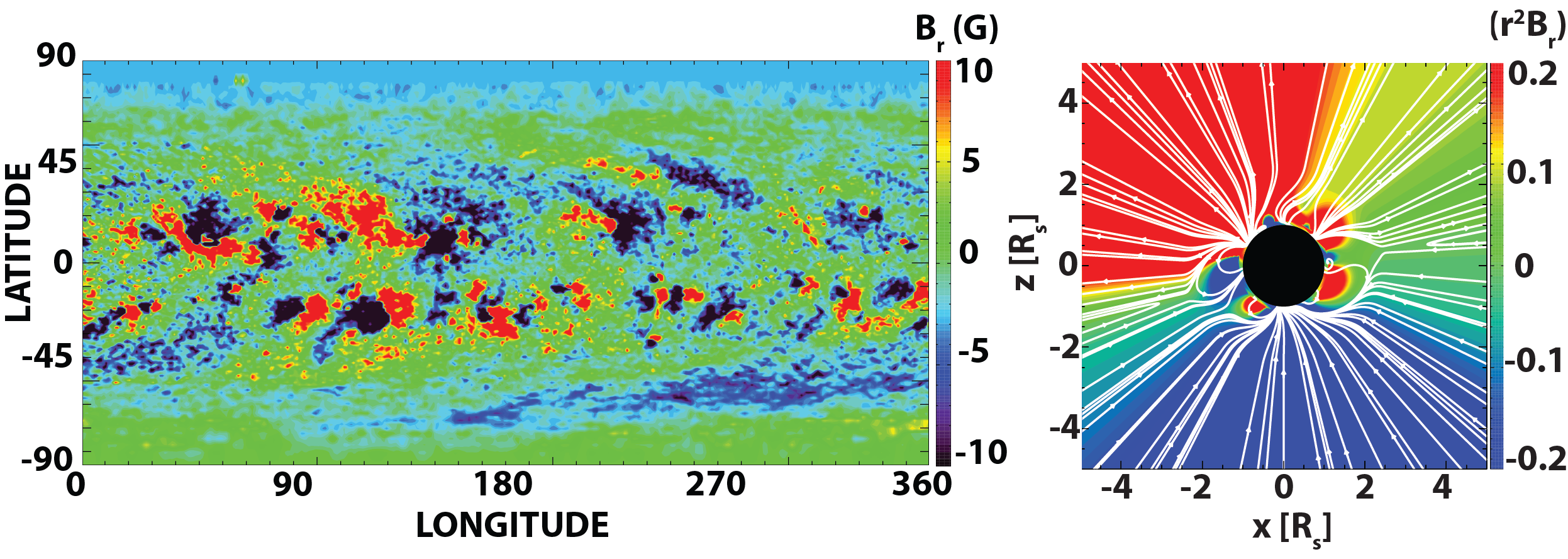}
\caption{Left: Radial magnetic field from GONG magnetogram observation at 2013 Apr 11, 06:04 UT. Right: Radial magnetic field calculated using PFSS Model with spherical harmonics of order 90. Simulations with 180 harmonics were less stable.}
\label{fig:gong}
\end{figure}

For comparison with the SA-MHD numerical result presented later in this paper, we provide numerical solution for the steady state solar corona prior to the SEP event occurred on  April 11, 2013 \cite[][]{Lario:2014a}. GONG magnetogram at 2013 Apr 11, 06:04 UT serves as the inner boundary condition and the main driver of the solar wind model (see the left panel of \figurename~\ref{fig:gong}).

In the simulation, the initial condition of the 3-D magnetic fields in the solar corona is reconstructed by a potential field source surface (PFSS) solution using a spherical harmonic expansion with the inner boundary provided by the magnetogram. The radial magnetic fields on the surface of the Sun calculated using PFSS is shown in the right panel of \figurename~\ref{fig:gong}. 
The PFSS reconstruction has been used to obtain the potential ($\bB_\snull$) field below the source surface located at $2.5R_\sSun$. 
Below the source surface we solve the full set of MHD equations (\ref{eq:continuity} and \ref{eq:inductionB0}-\ref{eq:totenergyB0}), above it we add the source term ${\bf S}_\sM$ given by \equationname~(\ref{eq:momentumsource}) to the momentum equation, (equation~\ref{eq:momentumB0}), as well as the extra energy source, $\bu\cdot{\bS}_\sM$, to the energy equation, (equation~\ref{eq:totenergyB0}). The governing equations for the solar corona and inner heliosphere models described by \cite{Sokolov:2021a} are solved using the SWMF \citep{Toth2005b,Toth:2012a} developed at the University of Michigan. 

\begin{figure*}[h!b]
\centering
\includegraphics[width=1\textwidth]{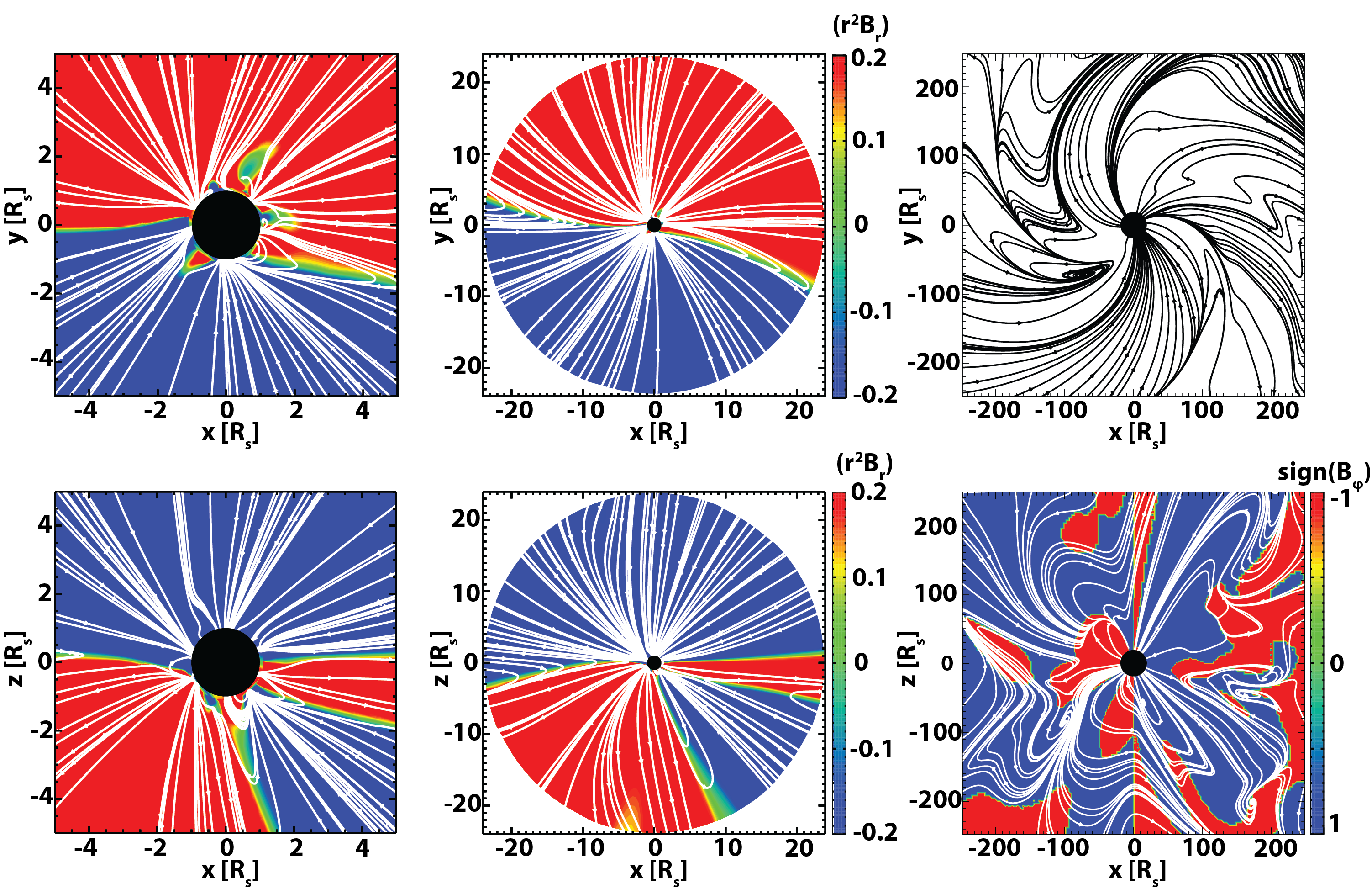}
{\caption{The magnetic field in the equatorial plane (top panels) and in the $y=0$ meridional planes (bottom panels) of the rotating HGR coordinate system. }
\label{fig:Bfull}}
\end{figure*}

The magnetic field in the equatorial plane (\figurename~\ref{fig:Bfull} top panels) and in the $y=0$ meridional plane of the rotating HGR coordinate system (\figurename~\ref{fig:Bfull} bottom panels) exhibits visible imperfections in the form of disconnected (``V-shaped'') field lines below 20$R_\sSun$ and too long closed loops in (the middle bottom panel of \figurename~\ref{fig:Bfull}). The magnetic field in the equatorial plane of the inner heliopshere (right panels in \figurename~\ref{fig:Bfull}) is totally disordered and is not applicable to solve magnetic connectivity. We emphasize that this feature is independent of the numerical solution algorithm: while steady-state field lines and stream-lines are identical at the PDE level, this is not true in any discretization within the framework of computational MHD. The stream-aligned model is designed to address all these issues.        

\subsection{The Leontovich Boundary Condition as the Aligning Source}
The computational SA-MHD cannot be applied to the entire solar corona that starts with a slowly expanding atmosphere with speeds as low as a few km/s on top of the transition region and in closed field regions. The slow-speed region extends to heliocentric distances of $2.5-3.5\,R_\sSun$. In this region the SA-MHD would fail to describe closed magnetic field lines. In addition, the problem to align the strong magnetic field with the slow and somewhat random motion is not only physically inappropriate, but also mathematically ill-posed, since the characteristic perturbation speeds tend to infinity if the plasma speed tends to zero. Therefore, we successfully apply the SA-MHD above some spherical heliocentric boundary with a radius of $R_{\sS\sA}$. 

Below this boundary we apply source terms in the MHD equation, which are proportional to the electric field describing the non-alignment between the field and flow and turning to zero, when the alignment is achieved. These terms reduce the electric field below $R_{\sS\sA}$ and entirely eliminate it at $R=R_{\sS\sA}$. However, they do not prevent slow motions and magnetic field line reconnection and closure. The sources are heuristically derived from the well-known \cite{Leontovich:1948a} boundary condition which make them physics-based.

The Leontovich boundary condition \cite[]{Leontovich:1948a, Landau:1985a} is applicable to  electromagnetic fields when a high-frequency wave interacts with a metal and deep inside the metal the field vanishes. The currents shielding the field are assumed to be concentrated in a thin layer near the boundary, having the impedance, $Z$. Under these assumptions the transverse components of electric and magnetic fields outside the metal are related to boundary condition:
\begin{equation}\label{eq:Leontovichbc}
\bE_t=Z \ \bB_t\times\bn,
\end{equation}
where $\bn$ is a unit vector normal to the metal surface directed inward the metal. A straightforward (however, insufficient) application to the boundary between the usual and SA-MHD gives us an opportunity to consider the electric-field-free SA-MHD region as if this is the metal and to assign the impedance $Z$ to the boundary surface. In this case at the MHD side of this boundary the finite electric field will cause a change in the $\bB_\snull$ field given by Eq.~(\ref{eq:Leontovichbc}):
\begin{equation}\label{eq:deltab0}
\delta\bB_\snull=\bn\times\bE/Z,
\end{equation}
It is important to emphasize that this field is induced by the current which is not conducted by the moving solar wind plasma, that is why \equationname~(\ref{eq:deltab0}) describes $\bB_\snull$ field, rather than $\bB_1$. Another comment is that the field undergoes a jump across the surface at which the finite current is concentrated, that is why the field given by \equationname~(\ref{eq:deltab0}) exists only ``outside the metal,'' that is in the SA-MHD region that starts at a heliocentric distance of $R=R_{\sS\sA}$.  
The modified field contributes to the flux function resulting in the evolution of the MHD state toward higher alignment (lower electric field). The effect is easy to evaluate if the impedance surface is orthogonal to the magnetic field. In this case the electric field, $\bE=B_n\bn\times\bu_t$ is orthogonal to $\bn$.  The normal components, $B_n=\bB\cdot\bn$, $u_n=\bu\cdot\bn$ of the field and velocity are not affected by the extra field given by Eq.~(\ref{eq:deltab0}), while the transverse velocity and field in the layer of a thickness of $\Delta x$, orthogonal to the magnetic field evolve since the extra field Eq.~(\ref{eq:deltab0}) is present at the upper boundary of the slab and absent at the lower one. Using the flux functions in equations (\ref{eq:inductionB0}) and (\ref{eq:momentumB0}), one can find the source terms for the MHD velocity, $(\partial_t\mathbf{u})_L$, and for the induction field, $(\partial_t\mathbf{B})_L$, caused by the shielding currents implied in the \cite{Leontovich:1948a} boundary condition  (that is why we use subscript $L$):
\begin{equation}
\left(\partial_t\bu\right)_L=\frac{B_n\delta\bB_\snull}{\rho\Delta x\muo}=-\frac{V_A^2\bu_t}{Z\Delta x}
\end{equation}
\begin{equation}
\left(\partial_t\bB\right)_L=-\frac{u_n\delta\bB_\snull}{\Delta x}=\frac{B_nu_n\bu_t}{Z\Delta x},
\end{equation}
in the last equation one can also put $B_nu_n=\bB\cdot\bu$, since the normal vector is aligned with the field, hence, $\bB=B_n\bn$. Now, one can derive {\it aligning sources} assuming the distributing conducting layers instead of concentrated impedance by substituting $\mathrm{d}(1/Z)/\mathrm{d}x$ for $1/(Z\Delta x)$ ratio, which may be conveniently parameterized as $\mathrm{d}(1/Z)/\mathrm{d}x=1/(V_A^2+u_n^2)\tau$, so that: 
\begin{equation}\label{eq:leonovichsourceu}
\left(\partial_t\bu\right)_L=-\frac{V_A^2\bu_t}{(V_A^2+u_n^2)\tau}
\end{equation}
\begin{equation}\label{eq:leonovichsourceB}
\left(\partial_t\bB\right)_L=\frac{(\bB\cdot\bu)\bu_t}{(V_A^2+u_n^2)\tau}.
\end{equation}
These sources at some choice of $\tau$ can be added to the MHD equations in a chosen region, not necessarily near the boundary. The advantage of the chosen parameterization is that $\tau$ directly characterizes the electric field relaxation due to the aligning sources, since:
\begin{equation}\label{eq:leonovichsourceE}
\left(\partial_t\bE\right)_L=B_n\bn\times\left(\partial_t\bu\right)_L-u_n\bn\times\left(\partial_t\bB\right)_L=-\frac\bE\tau.
\end{equation}
Although greatly improving the MHD solution at $R<R_{\sS\sA}$, the source terms do not allow us to align the solution at $R\rightarrow R_{\sS\sA}$. The latter goal is achieved if while integrating numerically the MHD equations over time after advancing the solution through the time step, $\Delta t$, we add the source terms and assume that the relaxation rate, $1/\tau$, is equal to $1/\Delta t$ at $R\rightarrow R_{\sS\sA}$ and modulated with some geometric factor, $\xi$, which equals one at $R \rightarrow R_{\sS\sA}$ and decays at smaller heights. In other words, the linear aligning operator is applied to the velocity and magnetic field vectors in the following manner:  
\begin{equation}\label{eq:leonovichoperatoru}
\bu\rightarrow\bu+\Delta\bu_L,\qquad\Delta\bu_L=-\frac{\xi V_A^2\bu_t}{V_A^2+u_n^2}
\end{equation}
\begin{equation}\label{eq:leonovichoperatorB}
\bB\rightarrow \bB+\Delta\bB_L,\qquad\Delta\bB_L=\frac{\xi(\bB\cdot\bu)\bu_t}{V_A^2+u_n^2}.
\end{equation}
The total energy density decreases by the following quantity:
\begin{equation}\label{eq:leonovichincrementE}
\Delta{\cal E}_L=\frac{\left(\frac{\xi^2}2-\xi\right)V_A^2\rho\bu^2_t}{V_A^2+u_n^2}.
\end{equation}

At $\xi=1$, that is in the proximity of the SA-MHD computational domain, operator \ref{eq:leonovichoperatoru}, \ref{eq:leonovichoperatorB} provides a perfect alignment about some weighted average direction such that the weak field is aligned with the strong stream without noticeable modification in the stream direction and, in the opposite limiting case, slow flow is aligned with the strong field.  

\subsection{Numerical Solution with SA-MHD}
The simulation results with SA-MHD are provided for the same solar configuration as described in Section 4.1. Below the PFSS source surface at $R=2.5\,R_\sSun$ the MHD model is not modified and the numerical solution is not much different from that provided in Fig.2.

\begin{figure*}[htb]
\centering
\includegraphics[width=1\textwidth]{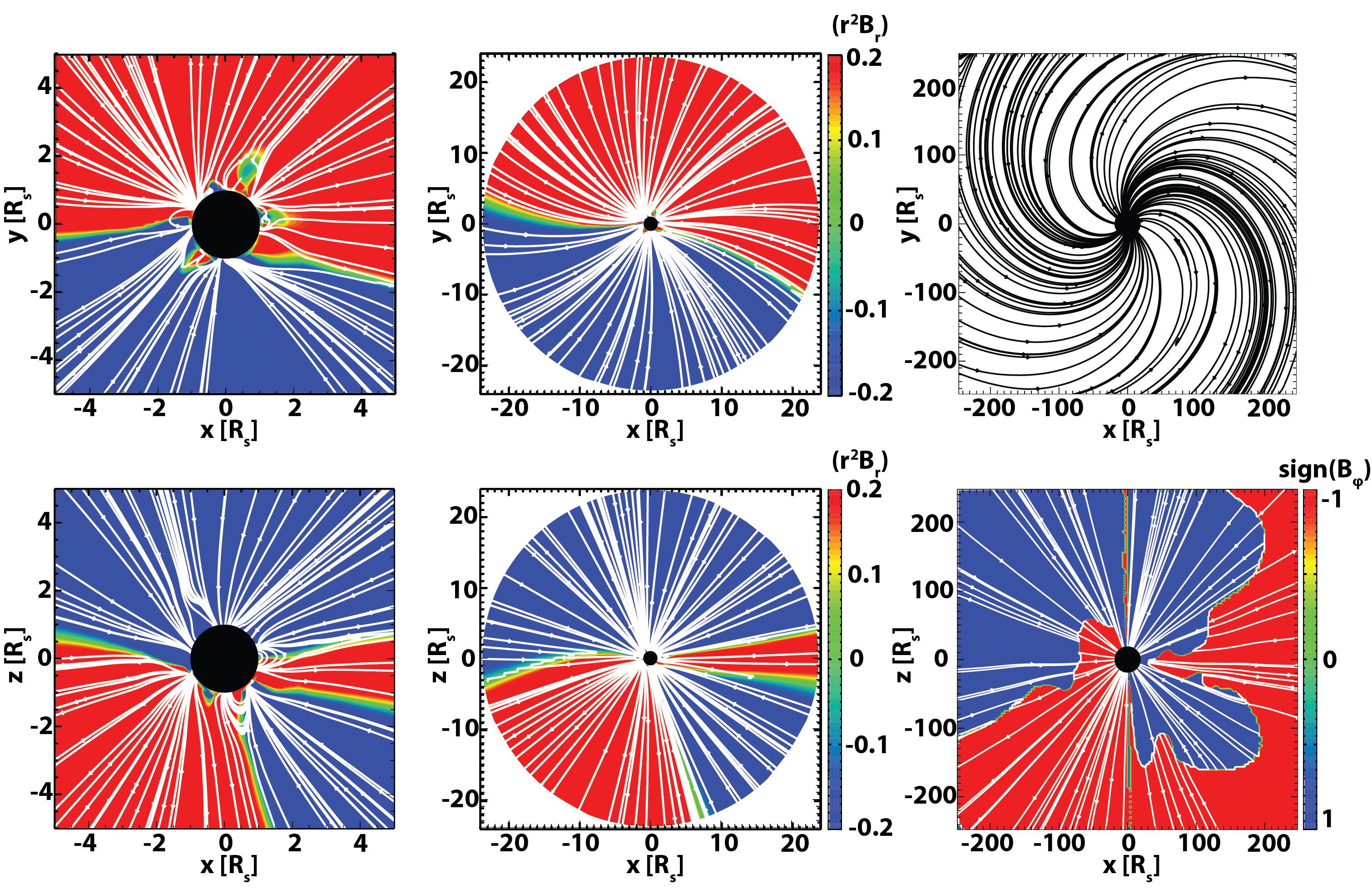}
{\caption{SA-MHD solutions of magnetic field lines in the equatorial plane of the inner heliosphere in the rotating HGR coordinate system.}
\label{fig:Bfull2}}
\end{figure*}

In the intermediate region $2.5\,R_\sSun<R<3.5\,R_\sSun$ we apply the aligning operator \ref{eq:leonovichoperatoru}, \ref{eq:leonovichoperatorB} based on the Leontovich boundary condition as described in Section 4.2. The geometric factor $\xi=R/R_\sSun-2.5$ is chosen equal to zero at the lower boundary and one at the upper boundary, of this region. 

At $R=R_{\rm SA}=3.5\,R_\sSun$ the MHD solution is enforced to be aligned, so that above this boundary the SA-MHD equations are solved numerically. The characteristic  perturbation speeds needed to construct the numerical scheme are discussed in Appendix.

\figurename~\ref{fig:Bfull2} shows the magnetic field structure in the corona and inner heliosphere obtained with SA-MHD and the \cite{Leontovich:1948a} boundary condition. One can see that the unphysical features seen in \figurename~\ref{fig:Bfull} are gone: There are no long closed loops or V-shaped field lines and the magnetic field lines in the inner heliosphere clearly form a \cite{Parker:1958a} spiral.

\subsection{Magnetic Connectivity}
\label{subsec:connectivity}
Magnetic connectivity (the magnetic field line connecting a point of interest in the heliosphere to the solar photosphere) is a very important component of space weather simulations. Among other things magnetic connectivity controls the simulated intensity and time profiles of SEP events \cite[\cf][]{Borovikov:2018a, GangLi:2021a, Young:2021a}.

\begin{figure*}[htb]
\centering
\includegraphics[width=0.475\textwidth]{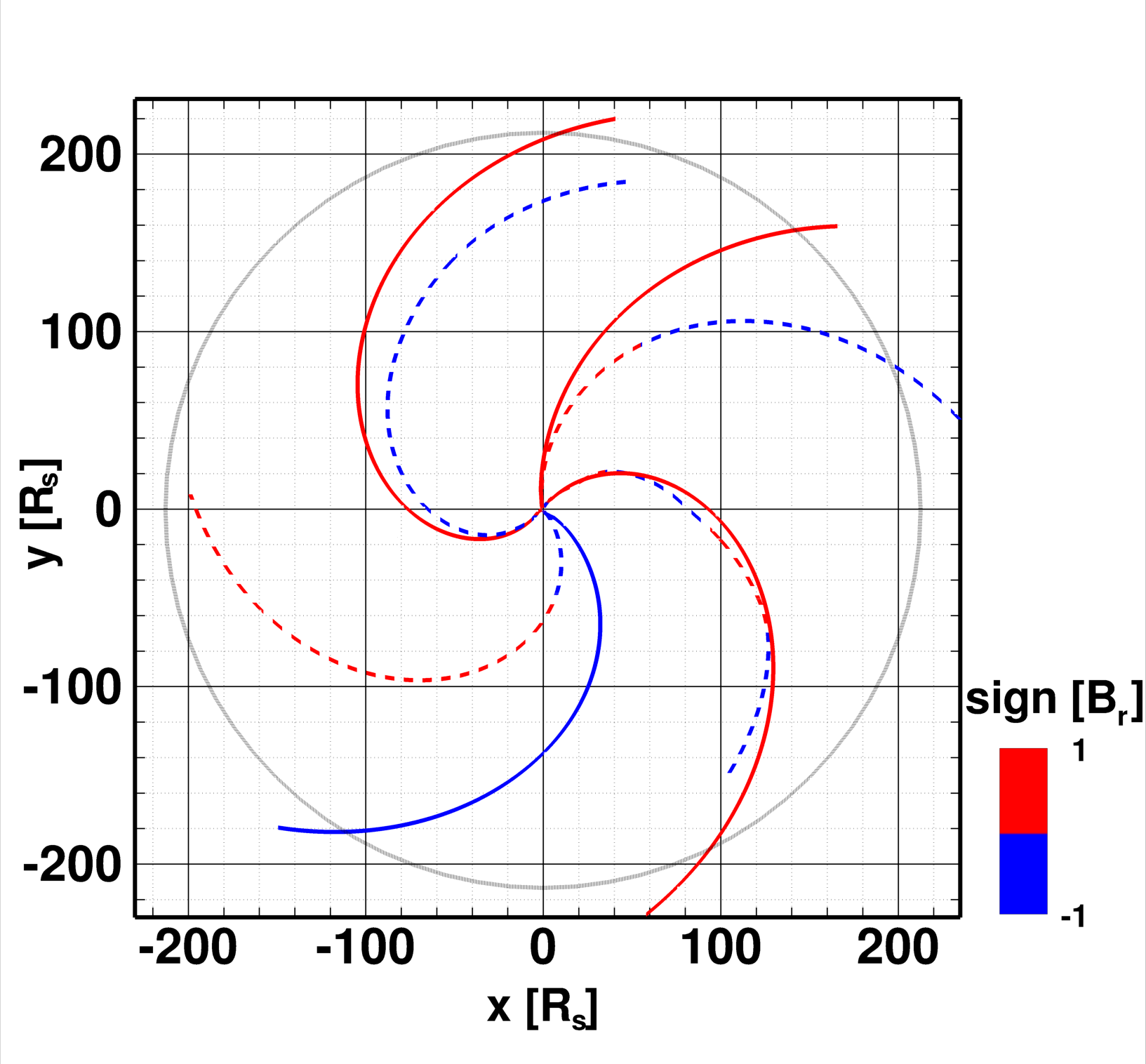}
{\caption{
Stream-lines obtained with MHD (dashed lines) and SA-MHD (solid lines). The stream-lines are colored by the sign of of the radial component of the local magnetic field vector. One can see that the sign of the radial magnetic field ``flips'' whenever the stream-line crosses the heliospheric current sheet.}
\label{fig:spiral}}
\end{figure*}

For the synoptic map shown in \figurename~\ref{fig:gong} the stream-lines are not as simple as in the case of uniform solar wind speed (see \figurename~\ref{fig:Parker}). As one see in 
\figurename~\ref{fig:spiral} the delay between the Earth crossing of MHD and SA-MHD stream-lines can vary between a few hours and a few days, depending on actual angular radial speed profile.

Most existing SEP models use MHD stream-lines as proxies for IMF lines and solve the SEP transport along these stream-lines. Our results indicate that this approach should be used with great care: the Earth crossing of an interplanetary stream-line can be off by a few hours or a few days depending on actual solar wind conditions. In addition, the sign of the radial magnetic field component can flip back and forth depending on the heliospheric current sheet crossings, so the $\mu=\cos(pitch-angle)$ quantity (appearing in the focused transport equation of \cite{Skilling1971a}) might abruptly change sign along the stream-line. An equation describing energetic particle transport in SA-MHD and a method to solve such equation numerically is described in a follow-up paper \citep{Sokolov:2021b}.

\textbf{\section{Conclusion}}
\label{sec:conclusion}
We showed how the MHD equations can be modified to describe stream-aligned steady-state flows of plasma parallel to the magnetic fields with no electric field. We call this approach stream-aligned MHD (SA-MHD). We applied the new approach to modelling the solar corona and heliosphere in a coordinate frame co-rotating with the Sun. Domains using the SA-MHD approach can gradually be merged to regions in which the ordinary MHD equations are solved by distributed external currents similar to those implied by the \cite{Leontovich:1948a} boundary condition. With this approach, the numerical solution in the heliosphere is perfectly aligned with the IMF lines shaped as classical Parker spirals. Our model addresses the magnetic connectivity problem as well as the simulation of the SEP acceleration and transport.

\begin{acknowledgments}
This work was supported by the NASA Heliophysics DRIVE Science Center (SOLSTICE) at the University of Michigan under grant NASA 80NSSC20K0600. We would also like to acknowledge high-performance computing support from: (1) Yellowstone (ark:/85065/d7wd3xhc) provided by NCAR's Computational and Information Systems Laboratory, sponsored by the National Science Foundation, and (2) Pleiades operated by NASA's Advanced Supercomputing Division.
\end{acknowledgments}
\appendix
\section{Characteristic Wave Speeds and Eigenvectors}
\label{app:waves}
In order to demonstrate an effect of the aligning terms on the characteristic waves, one can linearize Eqs.~(\ref{eq:aligningsource1}) and (\ref{eq:aligningsource2}), together with Eqs.~(\ref{eq:continuity}) and (\ref{eq:momentum}). For linear characteristic perturbations, $(\delta\rho,\, \delta\bu,\, \delta P,\,\delta\bB)^T$, propagating along the $x$ axis, i.e. depending on the coordinates and time via the combination, $(x-Dt)$. The background for the wave propagation is $(\rho_\snull,\, \bu_\snull,\, P_\snull,\, \bB_\snull)$ where $\bB_\snull = (B_{x\snull},B_{y\snull},0)$. The $z$ component of the magnetic field is made zero by choosing a coordinate system in which the magnetic field vector is in the $(x,y)$ plane. In the equations for the perturbation, all time derivatives should be substituted with $\partial_t = -D \partial_x$, and, after this substitution, one can omit the  $x$ derivatives  by denoting $\partial_x (\delta\rho) \rightarrow \delta\rho$ etc. Note, that \cite{DeSterck:1999} analyzed eigenvalues of two-dimensional steady-state SA-MHD, however, such eigenvalues only characterize the direction of wave propagation, rather than their speed needed to construct numerical schemes.

For the 8-wave MHD with the aligning sources the characteristic equations become:
\begin{subequations}
\begin{align}
    & -\Lambda \delta\rho + \rho_\snull\delta u_x = 0,
    \label{eq:deltarho}
\\
    & -\Lambda \delta u_x + \frac{\delta P}{\rho_\snull} + \frac{B_{y\snull}\delta B_y}{\muo\rho_\snull} = 0,
    \label{eq:deltaux}
\\
    & -\Lambda \delta u_y - \frac{B_{x\snull} \delta B_y}{\muo\rho_\snull} = 0,
    \label{eq:deltauy}
\\
    & -\Lambda \delta u_z - \frac{B_{x\snull} \delta B_z}{\muo\rho_\snull} = 0,
    \label{eq:deltauz}
\\
    & -\Lambda \delta P + \gamma P_\snull \delta u_x = 0,
    \label{eq:deltaP}
\\
    & -\Lambda \delta B_x + \alpha \left(\Lambda + u_{x\snull}\right) \delta u_x=0,
    \label{eq:deltaBx}
\\
    & -\Lambda \delta B_y + \left[\alpha \Lambda + \left(\alpha u_{x\snull} - B_{x\snull}\right) \right] \delta u_y + B_{y\snull} \delta u_x = 0,
    \label{eq:deltaBy}
\\
    & -\Lambda \delta B_z + \left[\alpha \Lambda + \left(\alpha u_{x\snull} - B_{x\snull}\right) \right] \delta u_z = 0,
    \label{eq:deltaBz}
\end{align}
\end{subequations}
where $\Lambda=D-u_{x\snull}$ is the characteristic speed in the plasma frame. 

Equations (\ref{eq:deltarho}) and (\ref{eq:deltaBx}) are independent and describe propagation of two perturbations with the plasma speed ($D=u_{x\snull},\,\Lambda=0$) in which
\begin{align}
    & (W1):
    \quad \Lambda_1 = 0, \quad \delta(\rho,u_x,u_y,u_z,P,B_x,B_y,B_z)^\sT= (1,0,0,0,0,0,0,0)^\sT \delta W_1,
    \label{eq:entropywave}
\\
    & (W6):
    \quad \Lambda_6 = 0, \quad \delta(\rho,u_x,u_y,u_z,P,B_x,B_y,B_z)^\sT= (0,0,0,0,0,1,0,0)^\sT \delta W_6.
    \label{eq:divBwave}
\end{align}
These are the entropy wave and the $\divg\bB$ wave respectively \cite[see][]{Powell:1999a}. Herewith, $\delta W_{1:8}$ denote arbitrary amplitudes of perturbations. 

Equations (\ref{eq:deltaBz}) and (\ref{eq:deltauz}) can be solved separately:
\begin{align}
&\Lambda_{5,7} \delta u_z= -\frac{V_{\sA,x}\delta B_z}{\sqrt{\muo\rho_\snull}},\quad V_{\sA,x}=\frac{B_{x\snull} }{\sqrt{\muo\rho_\snull}},\quad\tilde{V}_{\sA,x}=\frac{\alpha V_{\sA,x} }{\sqrt{\muo\rho_\snull}},\nonumber\\
&\Lambda_{5,7}=-\frac{\tilde{V}_{\sA,x}}{2}\mp\sqrt{\frac{\tilde{V}^2_{\sA,x}}{4}+V^2_{\sA,x}-u_{x\snull}\tilde{V}_{\sA,x}},
\label{eq:bothlimitingcases}.
\end{align}
This pair of waves can be considered in two limiting cases. In the first limiting case we consider ``traditional'' MHD with no alignment source. Now one can put $\alpha=0,\tilde{V}_{\sA,x}=0$ in \equationname~(\ref{eq:bothlimitingcases}). Note, that for $B_{x\snull}>0$ and $u_{x\snull}>0$ the perturbations of velocity and magnetic field are parallel in the left propagating wave ($\Lambda<0$), but they are anti-parallel in the wave propagating to the right, creating or increasing field misalignments with the stream. Both of these \alf waves can exist in the absence of an aligning source and propagate with the \alf speed, $V_{\sA,x}=B_{x\snull}/\sqrt{\muo\rho_\snull}$.

The other limiting case is when we include the aligning source and assume a stream aligned background ($B_{x\snull}=\alpha u_{x\snull}$). Under these circumstances,
$V^2_{\sA,x}=u_{x\snull}\tilde{V}_{\sA,x}$ and \equationname~(\ref{eq:bothlimitingcases}) describes two types of wave. In one of them,  propagating to the left (upwind), the perturbation is stream-aligned, $\delta B_z = \alpha\delta u_z$:
\begin{align}
\label{eq:w3}
    & (W5):
    \quad \Lambda_5= - \tilde{V}_{\sA,x}, 
    \quad \tilde{V}_{\sA,x} = 
    \frac{\alpha^2u_{x\snull}}{\muo\rho_\snull} = \frac{u_{x\snull}}{M_\sA^2}=\frac{V_{\sA,x}}{M_\sA}, \nonumber\\
    &\qquad\delta(\rho,u_x,u_y,u_z,P,B_x,B_y,B_z)^\sT= (0,0,0,1,0,0,0,\alpha)^\sT\delta W_5.
\end{align}
This means that the left \alf wave still exists, but its characteristic speed adjusts ($\tilde{V}_{\sA,x}\ne {V}_{\sA,x}$ at $M_\sA\ne1$) to ensure that the (parallel) perturbations of velocity and magnetic field are aligned. However, the right \alf wave in which the perturbations are anti-parallel (and they cannot be aligned) is arrested: \equationname \ref{eq:bothlimitingcases} requires the speed of this perturbation to be zero and the magnetic field to be unperturbed:  
\begin{align}
\label{eq:w7}
    & (W7):
    \quad \Lambda_7 = 0, \quad\delta(\rho,u_x,u_y,u_z,P,B_x,B_y,B_z)^\sT= (0,0,0,1,0,0,0,0)^\sT\delta W_7.
\end{align}

Analogously, equations (\ref{eq:deltaBy}) and (\ref{eq:deltauy}) allow a wave propagating with the plasma:
\begin{align}
\label{eq:w8}
    & (W8):
    \quad \Lambda_8=0,
    \quad\delta(\rho,u_x,u_y,u_z,P,B_x,B_y,B_z)^\sT= (0,0,1,0,0,0,0,0)^\sT\delta W_8.
\end{align}
Here, again, the perturbation speed is zero and the perturbations are not aligned. In the MHD description without the aligning sources this would be the slow magnetosonic wave. We see that in MHD (two 3-component-vector equations + two scalar equations, 8 characteristic waves) with the aligning sources and stream aligned background two kind of waves (right \alf and right slow magnetosonic) degenerate. These are the only waves in which the perturbations are not stream aligned. The distinction of SA-MHD (one 3-component-vector equation + three scalar equations with 6 characteristic waves) is that it {\it  lacks} these two degenerate waves since this approximation does not allow non-aligned perturbations. 

\subsection{Characteristic Waves in Stream Aligned MHD}
Instead of three components of magnetic field  we introduce a background value, $\alpha_\snull$,  and linear perturbation, $\delta\alpha$, for the quantity, $\alpha$, obeying the equation:
\begin{equation}\label{eq:deltaalpha}
-\Lambda\delta\alpha+\alpha_\snull\delta u_x = 0.
\end{equation}

In the entropy wave (W1) and the modified \alf wave (W5) we have $\delta\alpha=0$, while in the $\divg\bB$ wave (W6) the velocity perturbation should maintain the continuity of the stream-aligned field, $\delta\alpha u_{y\snull}+\alpha_\snull\delta u_y=0$, so that $\delta(\rho,u_x,u_y,u_z,P,\alpha)^\sT= (0,0,-u_{y\snull},0,0,\alpha_\snull)^\sT\delta W_6$.

In the remaining three branches of waves the magnetic field is stream-aligned and there is a relationship between $\delta u_y$ and $\delta B_y$:
\begin{equation}
\label{eq:w4a}
    \delta u_y = -\frac{\alpha_\snull u_{x\snull}}{\muo\rho_\snull\Lambda} \delta B_y, 
    \quad \left(\Lambda + \tilde{V}_{\sA,x}\right)\delta B_y - B_{y\snull}\delta u_x = 0.
\end{equation}
In the particular case of magnetic field aligned propagation ($B_{y\snull} = 0$) this reduces to the expression for another \alf wave of different polarization:
\begin{align}
\label{eq:w4b}
    & (W3a):
    \qquad \Lambda_3 = -\tilde{V}_{\sA,x},
    \quad\delta(\rho,u_x,u_y,u_z,P,\alpha)^\sT= (0,0,1,0,0,0)^\sT\delta W_2,
\end{align}
In the special case of $B_{y\snull} = 0$ equations (\ref{eq:deltaux}) and (\ref{eq:deltaP}) give two sound waves:
\begin{align}
\label{eq:w5a}
    & (W2a) \ \&\  (W4a):
    \quad \Lambda_{2,4}=\mp c_s,\quad c_s^2=\frac{\gamma P_\snull}{\rho_\snull},\nonumber\\
    & \delta(\rho,u_x,u_y,u_z,P,\alpha)^\sT= \left(\frac{\rho_\snull}{\Lambda_{2,4}},1,0,0,\frac{c_s^2\rho_\snull}{\Lambda_{2,4}},\frac{\alpha_\snull}{\Lambda_{2,4}}\right)^\sT\delta W_{2,4}.
    \hspace{5ex}
\end{align}
In the general case when $B_{y\snull} \ne 0$ there are three wave branches, in which the perturbations of velocity, pressure and magnetic field are all interrelated:
\begin{align}
\label{eq:w5b}
    & (W2b),\, (W3b) \ \&\  (W4b): \quad -\Lambda_{2:4}+\frac{c_s^2}{\Lambda_{2:4}} + \frac{V_{\sA,y}^2}{\Lambda_{2:4}+\tilde{V}_{\sA,x}}=0,
    \quad V_{\sA,y}^2 = \frac{B_{y\snull}^2}{\muo\rho_\snull},\\ 
    & \qquad 
    \delta(\rho,u_x,u_y,u_z,P,\alpha)^\sT= \left(\frac{\rho_\snull}{\Lambda_{2:4}},1,-\frac{\tilde{V}_{\sA,x} u_{y\snull}}
    {\Lambda_{2:4}\left(\Lambda_{2:4}+\tilde{V}_{\sA,x} \right)},0,\frac{c_s^2\rho_\snull}{\Lambda_{2:4}},\frac{\alpha_\snull}{\Lambda_{2:4}}\right)^\sT\delta W_{2:4}, \nonumber
\end{align}
This cubic equation always has three roots, corresponding to two fast and one slow magnetosonic wave. The characteristic speeds for $u_{\snull,x}>0$ range (from most negative to largest positive) as: left fast magnetosonic, left \alf, left slow magnetosonic, combined entropy$+\divgB$ wave, right fast magnetosonic wave. For $u_{\snull,x}<0$ the order is reversed, with changing ``left'' to ``right.''


\begin{thebibliography}{}
\expandafter\ifx\csname natexlab\endcsname\relax\def\natexlab#1{#1}\fi
\providecommand{\url}[1]{\href{#1}{#1}}
\providecommand{\dodoi}[1]{doi:~\href{http://doi.org/#1}{\nolinkurl{#1}}}
\providecommand{\doeprint}[1]{\href{http://ascl.net/#1}{\nolinkurl{http://ascl.net/#1}}}
\providecommand{\doarXiv}[1]{\href{https://arxiv.org/abs/#1}{\nolinkurl{https://arxiv.org/abs/#1}}}

\bibitem[{{Altschuler} {et~al.}(1977){Altschuler}, {Levine}, {Stix}, \&
  {Harvey}}]{alt77}
{Altschuler}, M.~D., {Levine}, R.~H., {Stix}, M., \& {Harvey}, J. 1977, Solar
  Physics, 51, 345

\bibitem[{Arge \& Pizzo(2000)}]{Arge:2000a}
Arge, C.~N., \& Pizzo, V.~J. 2000, J. Geophys. Res., 105, 10,465,
  \dodoi{10.1029/1999JA000262}

\bibitem[{{Biondo, Ruggero} {et~al.}(2021){Biondo, Ruggero}, {Bemporad,
  Alessandro}, {Mignone, Andrea}, \& {Reale, Fabio}}]{Biondo2021a}
{Biondo, Ruggero}, {Bemporad, Alessandro}, {Mignone, Andrea}, \& {Reale,
  Fabio}. 2021, J. Space Weather Space Clim., 11, 7,
  \dodoi{10.1051/swsc/2020072}

\bibitem[{{Borovikov} {et~al.}(2018){Borovikov}, {Sokolov}, {Roussev},
  {Taktakishvili}, \& {Gombosi}}]{Borovikov:2018a}
{Borovikov}, D., {Sokolov}, I.~V., {Roussev}, I.~I., {Taktakishvili}, A., \&
  {Gombosi}, T.~I. 2018, Astrophys. J., 864, 88,
  \dodoi{10.3847/1538-4357/aad68d}

\bibitem[{Cohen {et~al.}(2006)Cohen, Sokolov, Roussev, Arge, Manchester,
  Gombosi, Frazin, Park, Butala, Kamalabadi, \& Velli}]{Cohen:2007a}
Cohen, O., Sokolov, I.~V., Roussev, I.~I., {et~al.} 2006, Astrophys. J. Lett.,
  654, L163, \dodoi{10.1086/511154}

\bibitem[{{Cohen} {et~al.}(2007){Cohen}, {Sokolov}, {Roussev}, {Arge},
  {Manchester}, {Gombosi}, {Frazin}, {Park}, {Butala}, {Kamalabadi}, \&
  {Velli}}]{cohen07}
{Cohen}, O., {Sokolov}, I.~V., {Roussev}, I.~I., {et~al.} 2007, Astrophys. J.
  Lett., 654, L163, \dodoi{10.1086/511154}

\bibitem[{{Cranmer}(2010)}]{cran10}
{Cranmer}, S.~R. 2010, Astrophys. J., 710, 676,
  \dodoi{10.1088/0004-637X/710/1/676}

\bibitem[{{De Sterck} {et~al.}(1999){De Sterck}, Low, \&
  Poedts}]{DeSterck:1999}
{De Sterck}, H., Low, B.~C., \& Poedts, S. 1999, Physics of Plasmas, 6, 954,
  \dodoi{10.1063/1.873336}

\bibitem[{{Dmitruk} {et~al.}(2002){Dmitruk}, {Matthaeus}, {Milano}, {Oughton},
  {Zank}, \& {Mullan}}]{dmit02}
{Dmitruk}, P., {Matthaeus}, W.~H., {Milano}, L.~J., {et~al.} 2002, Astrophys
  J., 575, 571

\bibitem[{Godunov(1961)}]{Godunov:1961b}
Godunov, S.~K. 1961, Sov. Math. Dokl., 2, 947

\bibitem[{Godunov(1972)}]{Godunov:1972a}
---. 1972, in Numerical Methods for Mechanics of Continuum Medium, Vol.~1
  (Novosibirsk: Siberian Branch of USSR Acad. of Sci.), 26--34

\bibitem[{Gombosi {et~al.}(2021)Gombosi, Chen, Glocer, Huang, Jia, Liemohn,
  Manchester, Pulkkinen, Sachdeva, Shidi, Sokolov, Szente, Tenishev, Toth, {van
  der Holst}, Welling, Zhao, \& Zou}]{Gombosi:2021rev}
Gombosi, T., Chen, Y., Glocer, A., {et~al.} 2021, J. Space Weather Space Clim.,
  11, 41, \dodoi{10.1051/swsc/2021020}

\bibitem[{Gombosi {et~al.}(2017)Gombosi, Baker, Balogh, Erickson, Huba, \&
  Lanzerotti}]{Gombosi:2017a}
Gombosi, T.~I., Baker, D.~N., Balogh, A., {et~al.} 2017, Space Sci. Rev., 1,
  \dodoi{10.1007/s11214-017-0357-5}

\bibitem[{Gombosi {et~al.}(2002)Gombosi, Toth, {De~Zeeuw}, Hansen, Kabin, \&
  Powell}]{Gombosi:2002a}
Gombosi, T.~I., Toth, G., {De~Zeeuw}, D.~L., {et~al.} 2002, J. Comput. Phys.,
  177, 176, \dodoi{10.1006/jcph.2002.7009}

\bibitem[{Gosling(1993)}]{Gosling:1993a}
Gosling, J.~T. 1993, J. Geophys. Res., 98, 18937, \dodoi{10.1029/93JA01896}

\bibitem[{Harten {et~al.}(1983)Harten, Lax, \& {van Leer}}]{Harten:1983a}
Harten, A., Lax, P.~D., \& {van Leer}, B. 1983, SIAM Rev., 25, 35

\bibitem[{{Hu} {et~al.}(2000){Hu}, {Esser}, \& {Habbal}}]{hu00}
{Hu}, Y.~Q., {Esser}, R., \& {Habbal}, S.~R. 2000, J.~Geophys.~Res., 105, 5093

\bibitem[{Landau {et~al.}(1985)Landau, Lifshitz, \& Pitaevskii}]{Landau:1985a}
Landau, L.~D., Lifshitz, E.~M., \& Pitaevskii, L.~P. 1985, {E}lectrodynamics of
  {C}ontinuous {M}edia, 2nd edn. (Oxford, UK: Butterworth-Heinemann)

\bibitem[{Lario {et~al.}(2014)Lario, Raouafi, Zhang, Dresing, \&
  Riley}]{Lario:2014a}
Lario, D., Raouafi, N.~E., Zhang, J., Dresing, N., \& Riley, P. 2014,
  Astrophys. J., 8, 8, \dodoi{10.1088/0004-637X/797/1/8}

\bibitem[{Leontovich(1948)}]{Leontovich:1948a}
Leontovich, M. 1948, in Investigations on Propagation of Radio Waves, ed.
  B.~Vvedenskii (Moscow: AN SSSR), 5--10

\bibitem[{Li {et~al.}(2021)Li, Jin, Ding, Bruno, {de Nolfo}, Randol, Mays,
  Ryan, \& Lario}]{GangLi:2021a}
Li, G., Jin, M., Ding, Z., {et~al.} 2021, Astrophys. J., 919, 146,
  \dodoi{10.3847/1538-4357/ac0db9}

\bibitem[{{Li} \& {Habbal}(2003)}]{Habb03}
{Li}, X., \& {Habbal}, S.~R. 2003, Astrophys J. Lett, 598, L125

\bibitem[{{Lionello} {et~al.}(2014{\natexlab{a}}){Lionello}, {Velli}, {Downs},
  {Linker}, \& {Miki{\'c}}}]{Lionello:2014a}
{Lionello}, R., {Velli}, M., {Downs}, C., {Linker}, J.~A., \& {Miki{\'c}}, Z.
  2014{\natexlab{a}}, Astrophys. J., 796, 111,
  \dodoi{10.1088/0004-637X/796/2/111}

\bibitem[{{Lionello} {et~al.}(2014{\natexlab{b}}){Lionello}, {Velli}, {Downs},
  {Linker}, {Miki{\'c}}, \& {Verdini}}]{Lionello:2014b}
{Lionello}, R., {Velli}, M., {Downs}, C., {et~al.} 2014{\natexlab{b}},
  Astrophys. J., 784, 120, \dodoi{10.1088/0004-637X/784/2/120}

\bibitem[{Ogino \& Walker(1984)}]{Ogino:1984a}
Ogino, T., \& Walker, R.~J. 1984, Geophys. Res. Lett., 11, 1018

\bibitem[{Oran {et~al.}(2013)Oran, {van der Holst}, Landi, Jin, Sokolov, \&
  Gombosi}]{Oran:2013a}
Oran, R., {van der Holst}, B., Landi, E., {et~al.} 2013, Astrophys. J., 778,
  176, \dodoi{10.1088/0004-637X/778/2/176}

\bibitem[{Osman {et~al.}({2011})Osman, Matthaeus, Greco, \&
  Servidio}]{Osman:2011a}
Osman, K.~T., Matthaeus, W.~H., Greco, A., \& Servidio, S. {2011}, Astrophys.
  J. Lett., {727}, \dodoi{{10.1088/2041-8205/727/1/L11}}

\bibitem[{Parker(1958)}]{Parker:1958a}
Parker, E.~N. 1958, Astrophys. J., 128, 664, \dodoi{10.1086/146579}

\bibitem[{Pomoell \& Poedts(2018)}]{Pomoell:2018a}
Pomoell, J., \& Poedts, S. 2018, J. Space Weather Space Clim., 8, A35,
  \dodoi{10.1051/swsc/2018020}

\bibitem[{Powell {et~al.}(1999)Powell, Roe, Linde, Gombosi, \& {De
  Zeeuw}}]{Powell:1999a}
Powell, K., Roe, P., Linde, T., Gombosi, T., \& {De Zeeuw}, D.~L. 1999, J.
  Comput. Phys., 154, 284, \dodoi{10.1006/jcph.1999.6299}

\bibitem[{Roussev {et~al.}(2003)Roussev, Gombosi, Sokolov, Velli, Manchester,
  DeZeeuw, Liewer, Toth, \& Luhmann}]{Roussev:2003a}
Roussev, I.~I., Gombosi, T.~I., Sokolov, I.~V., {et~al.} 2003, Astrophys. J.,
  595, L57

\bibitem[{Shen {et~al.}(2018)Shen, Yang, Zhang, Wei, \& Feng}]{Shen:2018a}
Shen, F., Yang, Z., Zhang, J., Wei, W., \& Feng, X. 2018, Astrophys. J., 866,
  18

\bibitem[{Skilling(1971)}]{Skilling1971a}
Skilling, J. 1971, The Astrophysical Journal, 170, 265, \dodoi{10.1086/151210}

\bibitem[{Sokolov {et~al.}(2013)Sokolov, {van der Holst}, Oran, Downs, Roussev,
  Jin, Manchester, Evans, \& Gombosi}]{Sokolov:2013a}
Sokolov, I.~V., {van der Holst}, B., Oran, R., {et~al.} 2013, Astrophys. J.,
  764, 23, \dodoi{10.1088/0004-637X/764/1/23}

\bibitem[{Sokolov {et~al.}(2021a)Sokolov, {van der Holst}, Manchester, Ozturk,
  Szente, Taktakishvili, Tóth, Jin, \& Gombosi}]{Sokolov:2021a}
Sokolov, I.~V., {van der Holst}, B., Manchester, W.~B., {et~al.} 2021,
  Astrophys. J., 908, 172, \dodoi{10.3847/1538-4357/abc000}
\bibitem[{Sokolov {et~al.}(2021b)Sokolov, Sun, Tóth, Huang, Tenishev, Zhaao, Kota, Cohen \& Gombosi }]{Sokolov:2021b}
Sokolov, I.~V., {Sun}, H., {Tóth}, G., {et~al.} 2021,
  Journ. Comp. Phys. (submitted)

\bibitem[{{Suzuki} \& {Inutsuka}(2005)}]{Suzuki:2005a}
{Suzuki}, T.~K., \& {Inutsuka}, S.-i. 2005, Astrophys. J. Lett., 632, L49,
  \dodoi{10.1086/497536}

\bibitem[{Tanaka(1994)}]{Tanaka:1994a}
Tanaka, T. 1994, J. Comput. Phys., 111, 381

\bibitem[{{T{\'o}th} {et~al.}(2005){T{\'o}th}, {Sokolov}, {Gombosi}, {Chesney},
  {Clauer}, {de Zeeuw}, {Hansen}, {Kane}, {Manchester}, {Oehmke}, {Powell},
  {Ridley}, {Roussev}, {Stout}, {Volberg}, {Wolf}, {Sazykin}, {Chan}, {Yu}, \&
  {K{\'o}ta}}]{Toth2005b}
{T{\'o}th}, G., {Sokolov}, I.~V., {Gombosi}, T.~I., {et~al.} 2005, Journal of
  Geophysical Research (Space Physics), 110, A12226,
  \dodoi{10.1029/2005JA011126}

\bibitem[{Toth {et~al.}(2012)Toth, {van der Holst}, Sokolov, {De Zeeuw},
  Gombosi, Fang, Manchester, Meng, Najib, Powell, Stout, Glocer, Ma, \&
  Opher}]{Toth:2012a}
Toth, G., {van der Holst}, B., Sokolov, I.~V., {et~al.} 2012, J. Comput. Phys.,
  231, 870, \dodoi{10.1016/j.jcp.2011.02.006}

\bibitem[{{Tu} \& {Marsch}(1997)}]{tu97}
{Tu}, C.-Y., \& {Marsch}, E. 1997, Solar Phys., 171, 363

\bibitem[{Usmanov {et~al.}(2000)Usmanov, Goldstein, Besser, \&
  Fritzer}]{Usmanov:2000a}
Usmanov, A.~V., Goldstein, M.~L., Besser, B.~P., \& Fritzer, J.~M. 2000, J.
  Geophys. Res., 105, 12,675, \dodoi{10.1029/1999JA000233}

\bibitem[{Usmanov {et~al.}(2018)Usmanov, Matthaeus, Goldstein, \&
  Chhiber}]{Usmanov:2018a}
Usmanov, A.~V., Matthaeus, W.~H., Goldstein, M.~L., \& Chhiber, R. 2018, The
  Astrophysical Journal, 865, 25, \dodoi{10.3847/1538-4357/aad687}

\bibitem[{{van der Holst} {et~al.}(2014){van der Holst}, {Sokolov}, {Meng},
  {Jin}, {Manchester}, {Toth}, \& {Gombosi}}]{vanderHolst:2014a}
{van der Holst}, B., {Sokolov}, I.~V., {Meng}, X., {et~al.} 2014, Astrophys.
  J., 782, 81, \dodoi{10.1088/0004-637X/782/2/81}

\bibitem[{{Verdini} {et~al.}(2009){Verdini}, {Velli}, {Matthaeus}, {Oughton},
  \& {Dmitruk}}]{Verdini:2009a}
{Verdini}, A., {Velli}, M., {Matthaeus}, W.~H., {Oughton}, S., \& {Dmitruk}, P.
  2009, Astrophys. J. Lett., 708, L116, \dodoi{10.1088/2041-8205/708/2/l116}

\bibitem[{{Weber} \& {Davis}(1967)}]{Weber:1967a}
{Weber}, E.~J., \& {Davis}, Jr., L. 1967, Astrophys. J., 148, 217,
  \dodoi{10.1086/149138}

\bibitem[{Young {et~al.}(2021)Young, Schwadron, Gorby, Linker, Caplan, Downs,
  Török, Riley, Lionello, Titov, Mewaldt, \& Cohen}]{Young:2021a}
Young, M.~A., Schwadron, N.~A., Gorby, M., {et~al.} 2021, Astrophys. J., 909,
  160, \dodoi{10.3847/1538-4357/abdf5f}

\end{thebibliography}
\end{document}